\documentclass[aps,pre,onecolumn,superscriptaddress]{revtex4}
\usepackage[dvips]{graphicx}
\usepackage[dvips]{color}
\usepackage{breakurl,amsmath,amssymb,pst-node,eepic}
\usepackage{multirow}

\usepackage{subfigure}
\usepackage{tikz}
\usetikzlibrary{shapes, arrows, calc, positioning,matrix}
\usepackage{bbold}

\begin{document}

\title{Achieving competitive advantage in academia\\through early career coauthorship with top scientists}

\author{Weihua Li}
\affiliation{Department of Computer Science, University College London, 66-72 Gower Street, London WC1E 6EA, UK}
\affiliation{Systemic Risk Centre, London School of Economics and Political Sciences, Houghton Street, London WC2A 2AE, UK}

\author{Tomaso Aste}
\affiliation{Department of Computer Science, University College London, 66-72 Gower Street, London WC1E 6EA, UK}
\affiliation{Systemic Risk Centre, London School of Economics and Political Sciences, Houghton Street, London WC2A 2AE, UK}

\author{Fabio Caccioli}
\affiliation{Department of Computer Science, University College London, 66-72 Gower Street, London WC1E 6EA, UK}
\affiliation{Systemic Risk Centre, London School of Economics and Political Sciences, Houghton Street, London WC2A 2AE, UK}
\affiliation{London Mathematical Laboratory, 8 Margravine Gardens, London WC 8RH, UK}

\author{Giacomo Livan}
\email{g.livan@ucl.ac.uk}
\affiliation{Department of Computer Science, University College London, 66-72 Gower Street, London WC1E 6EA, UK}
\affiliation{Systemic Risk Centre, London School of Economics and Political Sciences, Houghton Street, London WC2A 2AE, UK}


\begin{abstract}
We quantify the long term impact that the coauthorship with established top-cited scientists has on the career of junior researchers in four different scientific disciplines. Through matched pair analysis, we find that junior researchers who coauthor work with top scientists enjoy a persistent competitive advantage throughout the rest of their careers with respect to peers with similar early career profiles. Such a competitive advantage materialises as a higher probability of repeatedly coauthoring work with top-cited scientists, and, ultimately, as a higher probability of becoming one. Notably, we find that the coauthorship with a top scientist has the strongest impact on the careers of junior researchers affiliated with less prestigious institutions. As a consequence, we argue that such institutions may hold vast amounts of untapped potential, which may be realised by improving access to top scientists.
\end{abstract}

\maketitle

\section{Introduction}

The availability of data about published research has led academia to increasingly study itself over the last few years. Ranging from the impact of publications \cite{hirsch2005index,lehmann2006measures} to grant attribution and staff hiring \cite{clauset2015systematic,clauset2017data}, all aspects of academic research and academic life are now studied quantitatively, as testified by the rise of the so called ``Science of Science'' \cite{zeng2017science,fortunato2018science}. 

Within this space, a number of studies have recently focused on scientific / academic impact \cite{wang2013quantifying,sinatra2016quantifying}. This, of course, requires an operational definition of impact. Although academic impact is a complex and multifaceted concept, nowadays it is increasingly equated to a scientist's ability to attract large numbers of citations. This, in turn, is mainly due to the fact that citations are reliably and consistently recorded across several disciplines, and, most importantly, to the fact that citation-based bibliometrics indicators are often used as metrics to rank scholars and determine their career advancements \cite{moher2018assessing}. For these reasons, the vast majority of studies in the literature use citation-based metrics as a proxy for academic impact \cite{zeng2017science}. 

A relevant theme within the devoted literature is that of identifying early indicators of long-lasting academic impact and their manifestation in a junior researcher's career (see, e.g., \cite{zeng2017science,acuna2012future}). This is a notoriously challenging task, since the aspects of a researcher's career which are easy to quantify do not necessarily yield a large predictive power. Indeed, the productivity of most scientists fluctuates heavily over time \cite{petersen2012persistence,way2017misleading}. Moreover, the unpredictability of the occurrence of a scientist's ``greatest hits'' over their career trajectory \cite{sinatra2016quantifying} complicates the matter even further.

Due to such challenges, a number of studies in recent years took a different approach by seeking to predict academic impact based on the \emph{visibility} of a junior researcher \cite{tahamtan2016factors,ale2014visibility}. Quantifying visibility presents its own challenges, as it encompasses a number of semi-qualitative aspects that contribute to provide a junior researcher's output with a competitive advantage with respect to output of the same quality published by peers with similar academic status and seniority. Factors that contribute to the visibility of a junior researcher are the following: ($i$) the journals where her research is published \cite{vieira2010citations,xia2011multiple}, ($ii$) the prestige of the institutions she and her coauthors are affiliated with \cite{amara2015can}, and ($iii$) the reputation of her more established coauthors \cite{petersen2014reputation} and, more generally, her academic social network \cite{sarigol2014predicting,contandriopoulos2016impact}.

The first two factors are somewhat easier to measure, thanks to the availability of multiple indices aimed at ranking journals (e.g., the impact factor \cite{garfield2006history}) and institutions \cite{liu2005academic}. This, in turn, translates into quantitative assessments of the long-term impact that such factors have on an academic career. For example, it has been shown that the academic prestige of the institution(s) a junior researcher is affiliated with correlates positively with long-term impact, as it leads to higher productivity \cite{way2019productivity} and a higher probability of securing a tenured position and, more generally, of ending up in a more influential position within a discipline \cite{bedeian2010doctoral,clauset2015systematic}. 

The third factor, i.e., the ``social factor'' that contributes to a researcher's visibility, is much harder to quantify, especially in the case of junior researchers, whose academic social network is still relatively sparse compared to that of established scientists. A number of studies have circumvented this problem by restricting a junior researcher's social network to her mentors and supervisors, showing in general that the supervision of an impactful mentor has beneficial effects on a prot\'eg\'e's academic career \cite{malmgren2010role,lienard2018intellectual}. In a similar spirit, a recent paper has revealed a ``chaperone effect'' \cite{sekara2018chaperone} in scientific publishing, showing that publishing in high-impact venues as a senior author is exceedingly more likely for scientists who have already done so in the past as junior researchers. 

The above body of work suggests that the protracted interaction between a junior researcher and a well established senior collaborator has long-lasting positive effects. In this paper, we take the above body of literature to its extreme consequences, and ask whether single events of interaction with top scientists can have career-altering effects on a junior researcher's future. Our main claim is that the mere coauthorship with a top scientist leads to a lasting competitive advantage in terms of impact. We demonstrate this by means of a matched pair experimental design, splitting a large pool of authors with long-lived academic careers into two groups - those who coauthored at least one paper with a top-cited scientist early on in their career and those who did not. We show that - all other things being equal - junior researchers belonging to the former group enjoy a persistent competitive advantage with respect to their peers belonging to the latter, which ultimately gives them a much better chance of becoming top-cited scientists themselves.  

In the following, we show the presence of such a competitive advantage for junior researchers across four different scientific disciplines, and we demonstrate the robustness of this finding after controlling for a number of potential confounding factors. Finally, we also show that this result yields significant predictive power, as it can be exploited to improve the predictability of a junior researcher's long term academic impact based on their early career indicators.

\section{Results}

Let us begin by introducing the operational definition of top scientist that we shall retain throughout the rest of the paper. We say that a researcher is a top scientist in a given year if she belongs to the top 5\% of cited authors in her discipline for that same year. Let us remark right away that in more than 95\% of cases in our dataset, once a researcher becomes a top scientist she remains one until the end of her career.

We begin our analysis by pooling together all researchers from four disciplines (Cell Biology, Chemistry, Physics, and Neuroscience - see Appendix \ref{app:data} and Appendix \ref{app:lists} for the lists of journals we consider for each discipline) whose career started between 1980 and 1998 and lasted at least 20 years, who have at least 10 publications, and who have published at least one paper every five years. In total, we have 22, 601 such researchers (see Table \ref{table:match} for a detailed breakdown in terms of disciplines). Within such pool of authors with long-lived careers, the unconditional probability of being a top scientist in the 20th career year is 24.8\%. Let us now proceed to condition this probability based on the institutional prestige a junior researcher is embedded in. We measure this by means of the average adjusted Nature Index (see Appendix \ref{app:prestige}) of the researcher's institution and of the institutions her coauthors are affiliated with.

\begin{figure*}[h!]
	\centering
	\includegraphics[width=.98\linewidth]{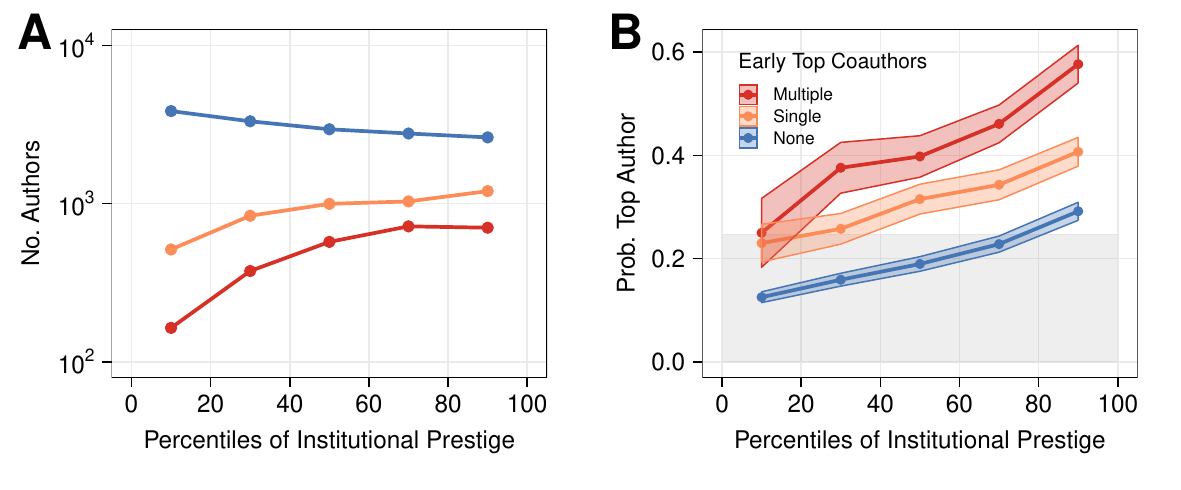}
	\caption{\textbf{Relationship between early career institutional prestige and probability of becoming a top scientist.} \textbf{A} Number of junior researchers in each quintile of the distribution of institutional prestige. \textbf{B} Probability of being a top scientist in the 20th career year as a function of institutional prestige (ribbon bands denote 95\% confidence intervals). In both panels authors are grouped based on whether in their first 3 career years they coauthored papers with one (orange), multiple (red) or no (blue) top scientists. The grey shaded area in panel \textbf{B} represents the unconditional probability of becoming a top scientist for the entire pool of junior researchers. 
	}\label{fig:prestige}
\end{figure*}

In Fig. \ref{fig:prestige} (left panel) we report the number of junior researchers falling within each quintile of the institutional prestige distribution, divided into three groups: those who did not coauthor a paper with a top-cited scientist early in their career (blue), those who coauthored papers with one top-cited scientist (orange), and those who coauthored papers with more than one top-cited scientist (red). In Fig. \ref{fig:prestige} (right panel) we show the probability for authors belonging to such groups of being a top-cited scientist themselves in their 20th career year.

The left panel reveals, as one would intuitively expect, a positive correlation between institutional prestige and coauthorsip with top scientists. The right panel, in turn, shows a positive correlation between institutional prestige and the probability of becoming a top-cited scientist in the long run. Yet, regardless of the relative position in terms of institutional prestige, such probability is significantly higher for researchers who coauthored papers with one top scientist, and markedly higher for those who did so with more than one top scientist.  

Furthermore, the right panel shows that, on average, the probability of becoming a top-cited scientist is below the aforementioned unconditional one (grey shaded area) for almost the entire pool of junior researchers lacking a top coauthor in their early career, with only those in the top quintile of institutional prestige managing to do better. Conversely, junior researchers who publish with top-cited scientists are in the opposite situation, and achieve better-than-average impact regardless of their position in terms of institutional prestige. 

We now proceed to expand this analysis by assessing how excellence in different aspects of academia relates with long term impact by splitting all junior researchers in our pool into 8 mutually exclusive groups based on early career performance according to different indicators. Namely, we consider institutional prestige (I), productivity (P), measured by the number of papers published within the first 3 career years, and the citations received within the first 3 career years (C). We group junior researchers depending on whether they belong to the top 10\% of authors across such dimensions\footnote{Authors are compared against their peers in the same discipline who started their career in the same year.}. For example, we label as I the group of researchers belonging to the top 10\% in terms of institutional prestige, as IP (IC) the group of researchers belonging to the top 10\% in both institutional prestige and productivity (citations), and as IPC the group of authors belonging to the top 10\% of all three dimensions.
 
\begin{figure*}[h!]
\vspace{-1cm}
	\centering
	\includegraphics[width=.9\linewidth]{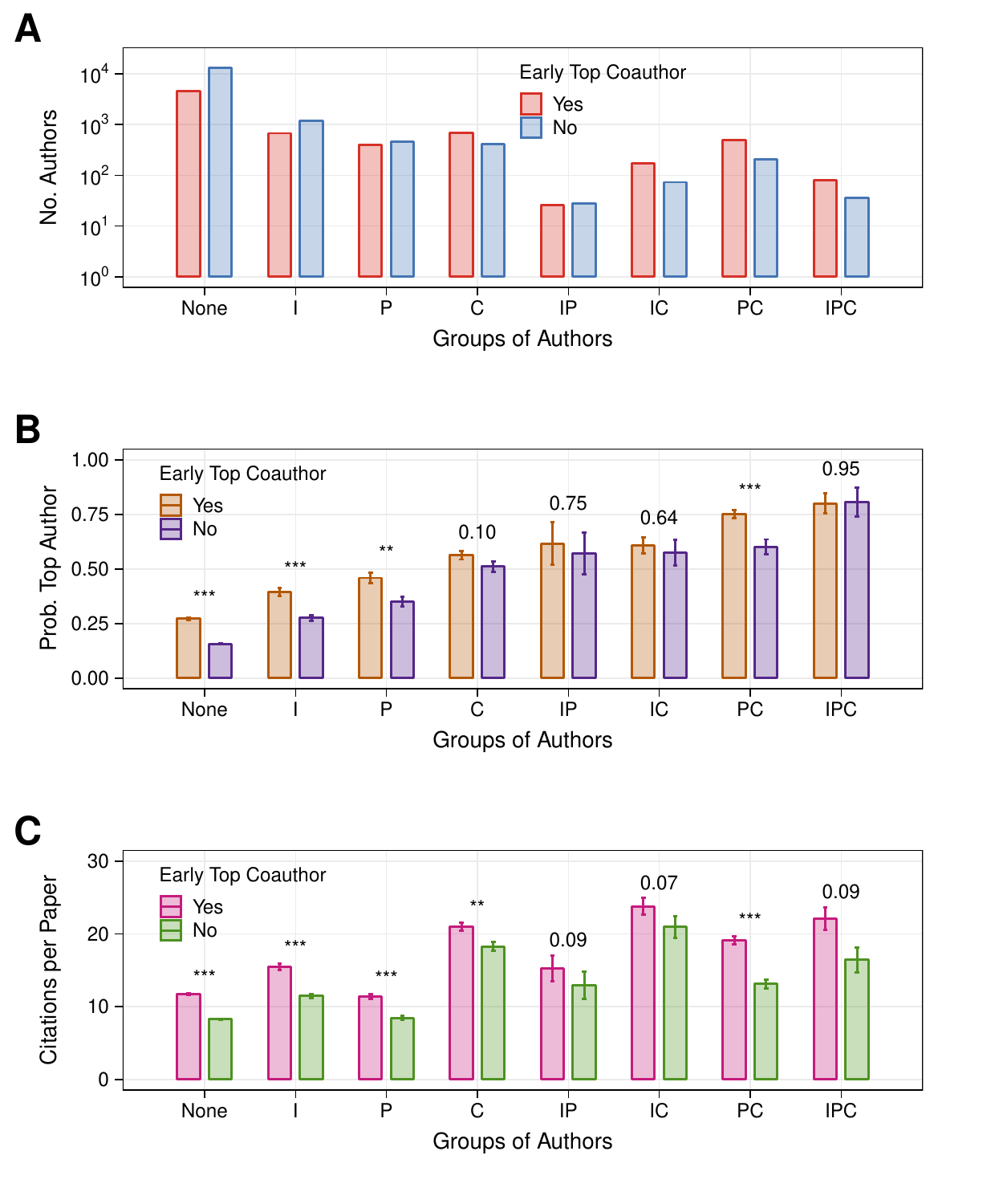}
	\caption{\textbf{Relationship between long term impact and early career performance.} \textbf{A} Number of junior researchers belonging to the top 10\% in various categories of early career performance (I denotes institutional prestige, P denotes productivity, C denotes citations received. All three such quantities are computed based on the first 3 career years). \textbf{B} Probability for authors belonging to each group of being a top scientist in their 20th career year. \textbf{C} Number of citations received per paper published by authors belonging to each group between their 4th and 20th career year. In panels \textbf{B} and \textbf{C} we report 95\% confidence intervals, and we report the $p$-values obtained via $t$-tests to assess the statistical significance of differences between the sub-group of junior researchers who coauthor work with a top scientist in the first 3 career years and the sub-group of those who do not. $^{*}$: $p<0.05$; $^{**}$: $p<0.01$; $^{***}$: $p <0.001$}
	\label{fig:pair}
\end{figure*}

The top panel in Fig. \ref{fig:pair} shows, for each category, the number of junior researchers in our dataset who coauthored at least one paper with a top-cited scientist vs. those who did not. As it can be seen, the only group where the latter are the clear majority is the one of researchers who do not belong to the top 10\% of any category. In all other cases, there is either a balance or a majority of junior researchers who coauthored with a top scientist, highlighting the presence of an overall positive correlation between coauthorship with top scientists and early career performance across all the dimensions we consider. 

The middle panel shows the probability of becoming a top-cited scientist for authors belonging to each of the above categories. Overall, independently of coauthorship with top scientists, we find this probability to be progressively higher as we consider authors belonging to the top 10\% of more categories, signalling a positive correlation between early and long-run career impact. Notably, such probability is above 50\% for junior researchers belonging to the top decile of two dimensions (IP, IC, PC), and hovers above 75\% for junior researchers in the top decile of all three categories (IPC).

However, in all categories except the latter we find the above probability to be systematically higher for the sub-groups of junior researchers with an early career paper coauthored with a top-cited scientist, and such differences are found to be statistically significant by a $t$-test in the cases labeled as ``None'', I, P, and PC. The results clearly show that the relative increase in the probability of becoming a top scientist tends to be larger in less exclusive groups, particularly in the group of junior researchers who do not belong to the top 10\% of any of the categories considered. Indeed, for this group the coauthorship with a top scientist almost doubles such probability, which jumps from 15.7\% to 27.2\%. Large increases in such probability are also apparent for the I and P groups. At the opposite end, coauthorship with a top scientist does not make a difference for junior researchers in the IPC group. One could interpret this as evidence that members of the latter group are with high probability already on the pathway to long-term career impact, regardless of their coauthors. In contrast, coauthorship with a top scientist truly has potential career-altering consequences for junior researchers who are not in the top 10\% of any of the categories we considered. In the following, we elaborate more on the mechanics leading to such consequences. 

The bottom panel in Fig. \ref{fig:pair} shows analogous results in terms of citations received per paper published between the 4th and 20th career year. We observe similar patterns to those shown in the middle panel, i.e., we find the sub-groups of junior researchers who coauthor with top scientists to systematically receive more citation than their peers in all categories. For the sake of readability, here we only show aggregate results. In Figures \ref{fig:cell_app}, \ref{fig:chemistry_app}, \ref{fig:neuroscience_app}, and \ref{fig:physics_app} (Appendix \ref{suppl}) we show equivalent figures for each of the four disciplines we consider. 

The above results begin to reveal a systematic competitive advantage for junior researchers who coauthor with a top scientist when considered as a \emph{group}, but do not yet quantify such advantage at the level of \emph{individual} careers. Figuratively speaking, this could only be measured by tracking a young researcher in two parallel careers where all factors remain identical, except that in one she gets to write a paper with a top scientist and in the other she does not. This is akin to a medical trial situation, where the effectiveness of a new drug has to be assessed by forming a treatment and a control group.  

We follow this line of reasoning and form two such groups in order to carry out a matched pair experimental design \cite{imai2009essential}. Namely, in each of the disciplines considered we identify pairs of junior researchers with similar early career profiles in terms of institutional prestige, productivity, and impact (i.e., number of citations accrued), with the only difference being that only one of the two has coauthored a paper with a top scientist during her first three career years (we shall refer to this as treatment). We then proceed to assess whether this has a detectable long-term effect by computing the average number of citations accrued between career years 4 and 20 by authors belonging to each group, both including and excluding those received by the papers published during the first 3 career years. In order to discount productivity as a possible confounding factor, we also compute the average number of citations received per paper published between career years 4 and 20. In particular, we focus on authors with low early career impact (i.e., with no more than 10 citations received in the first 3 career years) in order to focus on the group of junior researchers who can benefit the most from the interaction with a top scientist. Overall, there are 2,324 such authors in Cell Biology, 5,635 in Chemistry, 5,605 in Neuroscience and 5,414 in Physics.

\begin{table}[!ht]
\begin{center}
	\caption{\textbf{Matched pair analysis results.} Junior researchers are matched based on the institutional prestige they are embedded in, their productivity (measured by the number of papers published) and the number of citations received during their first 3 career years. One researcher per pair is either assigned to the treatment group (those who coauthored at least one paper with a top scientist) or the control group, and we compute the average of the following quantities across the two groups:  
		(A) Citations received in career years 1-20. 
		(B) Citations received in career years 4-20 excluding those received by the papers published in the first 3 career years.
	        (C) Citations received per paper published during career years 4-20.
	        (D) Number of different top scientists (per paper published) with whom the researcher has coauthored papers in career years 4-20 (excluding those already accounted in the first 3 career years for the treatment group).
	        (E) Number of times (per paper published) the researcher has coauthored papers with a top scientist in career years from 4 to 20 (excluding those already accounted in the first 3 career years for the treatment group).
	        ($^{*}$: $p<0.05$; $^{**}$: $p<0.01$; $^{***}$: $p <0.001$; NS: not significant). Numbers in brackets denote standard errors. }
	
	\hspace*{-2.3cm}
		\begin{tabular}{c|c|c|c|c|c|c|c|c|c|c|c|c}
		\hline \hline 
		\multirow{2}{*}{}  &  \multicolumn{3}{|c}{Cell Biology} &  \multicolumn{3}{|c}{Chemistry} &  \multicolumn{3}{|c}{Neuroscience} &  \multicolumn{3}{|c}{Physics} \\
		\hline

No. authors		       & \multicolumn{3}{|c}{$2,324$} & \multicolumn{3}{|c}{$5,635$}        & \multicolumn{3}{|c}{$5,605$} &\multicolumn{3}{|c}{$5,414$} \\
		
No. pairs		         & \multicolumn{3}{|c}{$468$} & \multicolumn{3}{|c}{$1,443$}      & \multicolumn{3}{|c}{$1,602$} &\multicolumn{3}{|c}{$1,362$}  \\
     	\hline 
		&  Treat & Control & $p$ &  Treat & Control & $p$ &  Treat & Control & $p$ &  Treat & Control & $p$  \\
		\hline 
  	 
Inst. prestige		    & $22.40$  & $22.24$   & NS     & $26.85$ & $26.99$ &  NS  		
             			& $10.52$ & $10.66$ &  NS 		& $19.58$  & $19.69$   & NS     \\
				        & $(0.57)$ & $(0.66)$  &        & $(0.32)$& $(0.39)$&  			
				        & $(0.17)$& $(0.19)$& 			& $(0.24)$ & $(0.27)$  &        \\
				        
Productivity			& $2.36$   & $2.37$    & NS     & $3.09$  & $3.13$  &  NS 		
						& $3.13$  & $3.09$  &  NS 		& $3.35$   & $3.25$    & NS     \\
				        & $(0.06)$ & $(0.06)$  &        & $(0.04)$& $(0.05)$&  			
				        & $(0.04)$& $(0.04)$& 			& $(0.05)$ & $(0.05)$  &        \\
				        				
Cit. (years 1-3)		& $4.80$   & $4.71$    & NS     & $4.06$  & $3.95$  &  NS 		
						& $3.95$  & $3.92$  &  NS 		& $4.22$   & $4.19$    & NS     \\
				        & $(0.15)$ & $(0.14)$  &        & $(0.08)$& $(0.08)$& 			
				        & $(0.07)$& $(0.07)$& 			& $(0.08)$ & $(0.08)$  &        \\

\hline	
\hline 
(A)						& $361.91$ & $281.74$&$^{***}$  & $301.36$& $246.38$&$^{***}$	
						& $383.13$ & $325.60$&$^{***}$ 	& $394.28$ & $288.31$  &$^{***}$\\
				        & $(18.70)$& $(15.13)$ &        & $(8.49)$& $(7.17)$&			
				        &$(10.68)$ &$(10.73)$& 			& $(13.88)$& $(9.47)$  &        \\

(B)	                    & $315.78$ & $247.25$&$^{***}$  & $259.68$& $209.86$&$^{***}$	
						& $314.56$& $273.18$&$^{**}$ 	& $343.09$ & $249.04$  &$^{***}$\\
				        & $(18.05)$& $(14.62)$ &        & $(8.18)$& $(6.80)$&			
				        &$(9.76)$&$(9.95)$& 			& $(13.05)$& $(8.92)$  &        \\

(C)	                    & $14.39$ & $11.56$ &$^{***}$   & $7.61$& $6.67$&$^{***}$		
						& $12.23$& $10.12$&$^{***}$ 	& $10.14$ & $8.59$  &$^{***}$\\
				        & $(0.48)$& $(0.44)$ &          & $(0.12)$& $(0.12)$& 			
				       	&$(0.24)$&$(0.20)$& 			& $(0.21)$& $(0.22)$ &        	\\

(D)	                    & $3.61$ & $2.72$  &$^{***}$      & $4.85$& $2.93$&$^{***}$ 		
						& $3.49$& $2.53$&$^{***}$ 		& $4.29$ & $2.68$  &$^{***}$\\
				        & $(0.17)$& $(0.16)$  &   	    & $(0.13)$& $(0.09)$&			
				        &$(0.09)$&$(0.08)$& 			& $(0.14)$& $(0.10)$  &        	\\

(E)	                    & $5.37$ & $4.05$  & $^{**}$       	& $10.71$& $5.97$&$^{***}$		
						& $6.91$& $4.98$&$^{***}$ 		& $9.73$ & $6.15$  &$^{***}$\\
				        & $(0.30)$& $(0.30)$ &        	& $(0.38)$& $(0.25)$&			
				        &$(0.24)$&$(0.22)$& 			& $(0.39)$& $(0.32)$  &        	\\

\hline \hline 
\end{tabular}
\label{table:match}
\end{center}
\end{table}		

The results of the analysis are reported in Table \ref{table:match}. In all four disciplines, we identify several hundreds of matched pairs of junior researchers with similar early career profiles, except for the presence / lack of a top coauthor. In all disciplines we find the treatment group of junior researchers who coauthored with a top scientist to achieve a higher impact, regardless of the specific citation metric, and we find the differences with respect to the control group to be statistically significant in all cases. 

This result demonstrates the long-lasting competitive advantage associated with early career coauthorship with top scientists. In order to understand the mechanism through which such a competitive advantage materialises, we measure how often on average junior researchers belonging to the two above groups get to coauthor papers with top-scientists between years 4 and 20 of their careers. The results from this analysis show that the treatment group consolidates its early competitive advantage by getting more opportunities to further collaborate with top scientists than the control group. This happens both in terms of the number of different top coauthors (excluding those already accounted for in the first 3 career years for the treatment group) and the number of individual coauthorship events with top scientists. Indeed, we find statistically significant differences between the treatment and control groups in all disciplines, with the former outperforming the latter in terms of repeated access to top scientists.

In Table \ref{table:match_app} (Appendix \ref{suppl}) we show that within pairs the junior researcher in the treatment group is the most cited in absolute terms ($p < .001$ in Chemistry, Physics, and Neuroscience, $p < .01$ in Cell Biology, one-tailed binomial test), and also the one who subsequently gets to coauthor more times with top scientists ($p < .001$ in Chemistry, Physics, and Neuroscience, $p = 0.38$ in Cell Biology, one-tailed binomial test).

\begin{figure*}[h!]
	\centering
	\includegraphics[width=.98\linewidth]{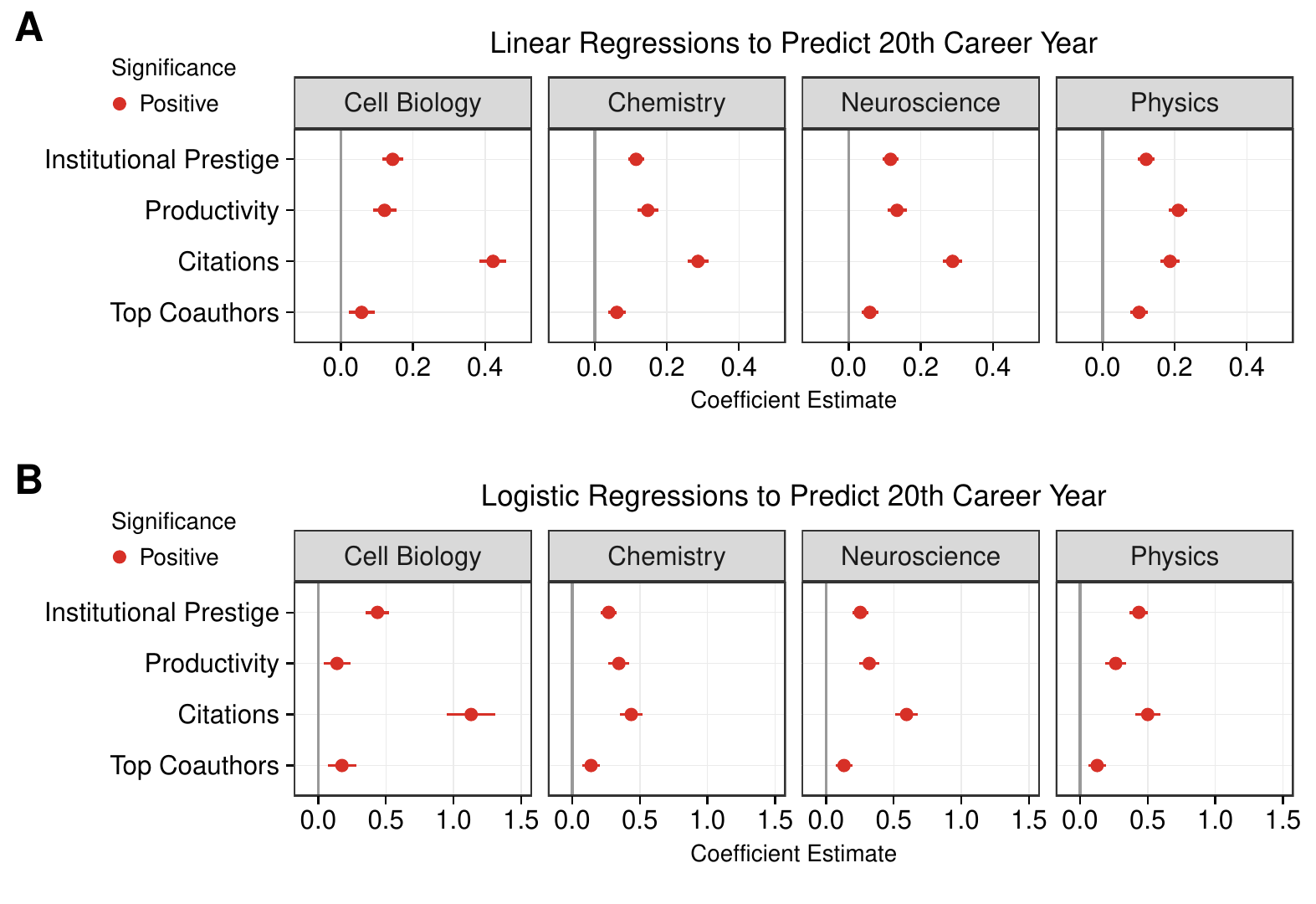}
	\caption{\textbf{Regression analysis of long-term career impact.} \textbf{A} Results form discipline-specific linear regressions whose dependent variable is the number of citations accrued in the first 20 career years ($R^2 = 0.33, 0.21, 0.18, 0.19$ for Cell Biology, Chemistry, Neuroscience, and Physics, respectively.) \textbf{B} Results from discipline-specific logistic regressions whose dependent variable is a binary indicator denoting whether a junior researcher is a top scientist in her 20th career year ($\mathrm{AUC} = 0.785, 0.711, 0.726, 0.749$ for Cell Biology, Chemistry, Neuroscience, and Physics, respectively). In all cases there is a statistically significant positive relationship between coauthorship with a top scientist in the first 3 career years and long term impact.}
	\label{fig:regressions}
\end{figure*}

Put together, the above results suggest that coauthorsip with a top scientist potentially represents a good predictor of impact in a long-lived academic career. This is confirmed by the outcomes of discipline-specific linear and logistic regressions, where we used early career coauthorship with at least one top scientist as a binary regressor against future impact, while controlling for institutional prestige, productivity, and impact in the first 3 career years (see the regression plot Fig. \ref{fig:regressions}). As dependent variables, we used the number of citations accrued in the first 20 career years in the case of linear regressions, and a binary variable to indicate whether a junior researcher had become a top scientist herself (i.e., among the top 5\% cited scientists in her discipline) in her 20th career year in the case of logistic regressions, respectively.

We systematically found coauthorship with a at least one top scientist to be a statistically significant predictor of long term future impact. Odds ratios for early collaboration with top coauthors in logistic regressions are: 1.19 for Cell Biology, 1.15 for Chemistry, 1.14 for Neuroscience, and 1.14 for Physics.

\section{Discussion}

In this paper we presented a number of analyses to assess the effect of early career coauthorship with established top-cited scientists on the long term prospects of junior researchers' academic impact. Invariably, our results highlighted that junior researchers who get the opportunity to coauthor at least one paper with a top scientist in the first three years of their career achieve a persistent competitive advantage with respect to their peers who do not get such an opportunity. Therefore, the following question becomes the crux of the matter: is such a competitive advantage a reflection of a young researcher's exceptional skills, which in turn lead her to collaborate with a top scientist, or is it instead a direct consequence of the interaction with a top scientist? 

Our results cannot provide a definitive answer to the above question for one fundamental reason, i.e., that we cannot control for the fact that established top scientists might attract the very best students. This acts as an ineradicable confounding factor in our analysis. Nevertheless, our results are systematic enough to suggest that it is the collaboration with a top scientist that creates the aforementioned competitive advantage, whose echo can still be detected 20 years later.  

Let us clarify that our results do not imply that all successful careers are launched thanks to the interaction with a top scientist early on. In fact, the probability to become a top scientist in the long run is the highest for those junior researchers who start their careers as the best among their peers (see Fig. \ref{fig:pair}). Nevertheless, for those junior researchers who are not at the top among their peers in at least one category among institutional prestige, productivity and impact, the opportunity to coauthor papers with top scientists systematically provides a competitive advantage. Loosely speaking, top junior researchers will always tend to become top scientists, whereas others may genuinely experience a shift in career trajectory after the interaction with a top scientist. In this respect, our work sheds light on previously published studies on the interactions between junior and well established scientists \cite{amjad2017standing,qi2017standing}.

The aforementioned competitive advantage materialises by means of a ``rich-get-richer'' mechanism, where the early career opportunity to coauthor papers with a top scientist translates into a higher probability of doing it again at later career stages, and, eventually, to become one. In this respect, our results are in line with other studies that have shown how academic achievements facilitate access to further impact and recognition \cite{petersen2011quantitative,way2019productivity}.

The present work sheds new light on the determinants of academic impact. Indeed, our results show that early career opportunities can play an exceedingly important role in shaping the prospects of a long academic career. Loosely speaking, we may say that being ``in the right place at the right time'' - and being able to seize on the opportunity - provides a junior researcher with an early edge which may separate her from her peers for years to come. This is highlighted very clearly by our matched pair analysis (see Table \ref{table:match}), which shows that the interaction with a top scientist in the first 3 career years is already enough to permanently split career trajectories that were otherwise on the same path. We interpret this result as evidence that academic institutions are filled with untapped potential, which may largely remain unrealised simply due to a lack of opportunity.

Seen from a different perspective, the above considerations suggest that top scientist can make a unique difference towards unlocking growth potential, especially for junior researchers in less prestigious institutions, who in fact are the most positively impacted when presented with the opportunity to work with a top scientist (see Fig. \ref{fig:pair}).

\bibliography{scibib}
\newpage

\appendix

\section{Data}
\label{app:data}

We collected publication and citation data for four disciplines (Cell Biology, Chemistry, Neuroscience, and Physics) indexed on the Web of Science database. For each discipline, we collected data about all papers published since 1970 in a selection of journals, their authors and their affiliations. The data include outputs such as letters and editorials, but we limited our dataset to standard articles and review articles, as these are the usual outputs of research efforts.

We selected journals based on two criteria: in the case of Chemistry and Physics, we selected all publications issued by the American Physical Society (APS, 9 journals) and the American Chemical Society (ACS, 42 journals), which represent major publishers in their respective fields. In the case of Cell Biology and Neuroscience, instead, we selected all journals whose publications collectively accrued at least 10,000 total citations according to the Journal Citation Reports (JCR). These amount to 53 journals in Cell Biology and 59 journals in Neuroscience, respectively. We provide the full lists of journals in Appendix \ref{app:lists}.

We then proceeded to disambiguate the names of the authors of papers in the above venues with the methodology published in \cite{sinatra2016quantifying}, and we only retained the papers and citations belonging to authors with at least 10 citations in their final career year. With these positions, we retained 226,362 papers and 71,794 authors in Cell Biology, 524,639 papers and 123,513 authors in Chemistry, 395,246 papers and 102,074 authors in Neuroscience, 412,063 papers and 80,218 authors in Physics.

\section{Institutional prestige score}
\label{app:prestige}

We measure the institutional prestige a junior researcher is embedded in by means of the Nature Index (https://www.natureindex.com/), which has been introduced by the Nature group in order to rank the academic prestige of universities and research institutions. This is computed by counting the number of papers published in a set of expert-selected journals in the above four disciplines (Cell Biology, Chemistry, Neuroscience and Physics). We adopt the same methodology and compute a given institution $i$'s prestige score as $\sqrt{P^\mathrm{nat}_i}$, where $P^\mathrm{nat}_i$ is the number of publications authored by researchers affiliated with institution $i$ in the Nature Index's list of journals since 1970. We then compute a paper's prestige score as the average prestige score of its authors' institutions, and a researcher's prestige score as the average prestige score of her papers.

\newpage
\section{Lists of journals}\
\label{app:lists}

\textbf{Journals in Cell Biology} \\
AMERICAN JOURNAL OF PHYSIOLOGY-CELL PHYSIOLOGY  \\
AMERICAN JOURNAL OF RESPIRATORY CELL AND MOLECULAR BIOLOGY  \\
AUTOPHAGY  \\
BIOCHIMICA ET BIOPHYSICA ACTA-MOLECULAR CELL RESEARCH  \\
CANCER CELL                        \\                                
CELL  \\
CELL AND TISSUE RESEARCH  \\
CELL CYCLE  \\
CELL DEATH \& DISEASE  \\
CELL DEATH AND DIFFERENTIATION  \\
CELL METABOLISM  \\
CELL REPORTS  \\
CELL RESEARCH  \\
CELL STEM CELL  \\
CELLULAR AND MOLECULAR LIFE SCIENCES  \\
CELLULAR PHYSIOLOGY AND BIOCHEMISTRY  \\
CELLULAR SIGNALLING  \\
COLD SPRING HARBOR PERSPECTIVES IN BIOLOGY  \\
CURRENT BIOLOGY  \\
CURRENT OPINION IN CELL BIOLOGY  \\
CURRENT OPINION IN STRUCTURAL BIOLOGY  \\
DEVELOPMENTAL CELL  \\
EMBO JOURNAL  \\
EMBO REPORTS  \\
EXPERIMENTAL CELL RESEARCH  \\
FASEB JOURNAL  \\
FEBS LETTERS  \\
GENES \& DEVELOPMENT  \\
INTERNATIONAL JOURNAL OF BIOCHEMISTRY \& CELL BIOLOGY  \\
JOURNAL OF CELL BIOLOGY  \\
JOURNAL OF CELL SCIENCE  \\
JOURNAL OF CELLULAR AND MOLECULAR MEDICINE \\
JOURNAL OF CELLULAR BIOCHEMISTRY \\ 
JOURNAL OF CELLULAR PHYSIOLOGY  \\
JOURNAL OF LEUKOCYTE BIOLOGY  \\
JOURNAL OF MOLECULAR AND CELLULAR CARDIOLOGY  \\
MOLECULAR AND CELLULAR BIOCHEMISTRY  \\
MOLECULAR AND CELLULAR BIOLOGY  \\
MOLECULAR AND CELLULAR ENDOCRINOLOGY  \\
MOLECULAR BIOLOGY OF THE CELL  \\
MOLECULAR CELL  \\
NATURE CELL BIOLOGY  \\
NATURE MEDICINE  \\
NATURE REVIEWS MOLECULAR CELL BIOLOGY  \\
NATURE STRUCTURAL \& MOLECULAR BIOLOGY  \\
ONCOGENE  \\
PLANT AND CELL PHYSIOLOGY  \\
PLANT CELL CIENCE SIGNALING  \\
SCIENCE TRANSLATIONAL MEDICINE  \\
STEM CELLS  \\
STRUCTURE  \\
TRENDS IN CELL BIOLOGY \\ \\

\textbf{Journals in Chemistry (published by American Chemical Society)} \\
ACCOUNTS OF CHEMICAL RESEARCH \\
ACS APPLIED MATERIALS \& INTERFACES  \\
ACS BIOMATERIALS SCIENCE \& ENGINEERING  \\
ACS CATALYSIS  \\
ACS CHEMICAL BIOLOGY  \\
ACS CHEMICAL NEUROSCIENCE  \\
ACS COMBINATORIAL SCIENCE   \\
ACS INFECTIOUS DISEASES   \\
ACS MACRO LETTERS  \\
ACS MEDICINAL CHEMISTRY LETTERS  \\
ACS NANO   \\
ACS PHOTONICS   \\
ACS SUSTAINABLE CHEMISTRY \& ENGINEERING  \\
ACS SYNTHETIC BIOLOGY  \\
ANALYTICAL CHEMISTRY  \\ 
BIOCHEMISTRY  \\
BIOCONJUGATE CHEMISTRY   \\
BIOMACROMOLECULES   \\
BIOTECHNOLOGY PROGRESS  \\
CHEMICAL RESEARCH IN TOXICOLOGY  \\
CHEMISTRY OF MATERIALS  \\
CRYSTAL GROWTH \& DESIGN   \\
ENERGY \& FUELS   \\
ENVIRONMENTAL SCIENCE \& TECHNOLOGY  \\
ENVIRONMENTAL SCIENCE \& TECHNOLOGY LETTERS  \\
INDUSTRIAL \& ENGINEERING CHEMISTRY RESEARCH   \\
INORGANIC CHEMISTRY  \\
JOURNAL OF AGRICULTURAL AND FOOD CHEMISTRY  \\
JOURNAL OF CHEMICAL EDUCATION  \\
JOURNAL OF CHEMICAL INFORMATION AND MODELING  \\
JOURNAL OF CHEMICAL THEORY AND COMPUTATION   \\
JOURNAL OF MEDICINAL CHEMISTRY  \\
JOURNAL OF NATURAL PRODUCTS  \\
JOURNAL OF PROTEOME RESEARCH  \\
JOURNAL OF THE AMERICAN CHEMICAL SOCIETY   \\
LANGMUIR  \\
MACROMOLECULES \\
MOLECULAR PHARMACEUTICS  \\
NANO LETTERS  \\
ORGANIC LETTERS  \\
ORGANIC PROCESS RESEARCH \& DEVELOPMENT  \\
ORGANOMETALLICS  \\

\textbf{Journals in Neuroscience} \\
ACTA NEUROPATHOLOGICA \\
ANNALS OF NEUROLOGY  \\
BEHAVIOURAL BRAIN RESEARCH  \\
BIOLOGICAL PSYCHIATRY  \\
BRAIN  \\
BRAIN BEHAVIOR AND IMMUNITY  \\
BRAIN RESEARCH  \\
CEREBRAL CORTEX  \\
CLINICAL NEUROPHYSIOLOGY  \\
CURRENT OPINION IN NEUROBIOLOGY \\
EUROPEAN JOURNAL OF NEUROLOGY  \\
EUROPEAN JOURNAL OF NEUROSCIENCE  \\
EXPERIMENTAL BRAIN RESEARCH  \\
EXPERIMENTAL NEUROLOGY \\
FRONTIERS IN HUMAN NEUROSCIENCE  \\
GAIT \& POSTURE  \\
GLIA  \\
HUMAN BRAIN MAPPING  \\
JOURNAL OF ALZHEIMERS DISEASE \\
JOURNAL OF CEREBRAL BLOOD FLOW AND METABOLISM  \\
JOURNAL OF COGNITIVE NEUROSCIENCE  \\
JOURNAL OF COMPARATIVE NEUROLOGY  \\
JOURNAL OF NEUROCHEMISTRY  \\
JOURNAL OF NEUROIMMUNOLOGY  \\
JOURNAL OF NEUROPHYSIOLOGY  \\
JOURNAL OF NEUROSCIENCE \\
JOURNAL OF NEUROSCIENCE METHODS  \\
JOURNAL OF NEUROSCIENCE RESEARCH  \\
JOURNAL OF NEUROTRAUMA  \\
JOURNAL OF PHYSIOLOGY-LONDON  \\
JOURNAL OF THE NEUROLOGICAL SCIENCES  \\
MOLECULAR NEUROBIOLOGY  \\
MOLECULAR PSYCHIATRY  \\
MULTIPLE SCLEROSIS JOURNAL  \\
MUSCLE \& NERVE  \\
NATURE NEUROSCIENCE  \\
NEURAL COMPUTATION  \\
NEURAL NETWORKS  \\
NEUROBIOLOGY OF AGING  \\
NEUROBIOLOGY OF DISEASE  \\
NEUROIMAGE  \\
NEURON  \\
NEUROPHARMACOLOGY  \\
NEUROPSYCHOLOGIA  \\
NEUROPSYCHOPHARMACOLOGY  \\
NEUROREPORT  \\
NEUROSCIENCE  \\
NEUROSCIENCE AND BIOBEHAVIORAL REVIEWS  \\
NEUROSCIENCE LETTERS  \\
PAIN  \\
PHARMACOLOGY BIOCHEMISTRY AND BEHAVIOR  \\
PROGRESS IN NEUROBIOLOGY  \\
PSYCHONEUROENDOCRINOLOGY  \\
PSYCHOPHARMACOLOGY  \\
PSYCHOPHYSIOLOGY  \\
SLEEP  \\
TRENDS IN COGNITIVE SCIENCES  \\
TRENDS IN NEUROSCIENCES  \\
VISION RESEARCH  \\
	
\textbf{Journals in Physics} \\
PHYSICAL REVIEW A \\
PHYSICAL REVIEW ACCELERATORS AND BEAMS  \\
PHYSICAL REVIEW APPLIED  \\
PHYSICAL REVIEW B \\
PHYSICAL REVIEW C                      \\ 
PHYSICAL REVIEW D  \\
PHYSICAL REVIEW E                 \\
PHYSICAL REVIEW LETTERS \\
PHYSICAL REVIEW X  \\

\section{Supplementary tables and figures}
\label{suppl}

\begin{table}[ht!]
	\centering
	\caption{\textbf{Binomial tests for matched pairs of junior researchers.} Junior researchers are matched based on the institutional prestige they are embedded in, the citations received and the number of papers published during their first 3 career years. One researcher per pair is either assigned to the treatment group (those who coauthored at least one paper with a top scientist). We compute the percentage of pairs in which the junior researcher who coauthored with a top scientist had a better performance with respect to the following quantities (statistical significance refers to a one-tailed binomial test): 	
		(A) citations received in career years 1-20;
		(B) citations received in career years 4-20 excluding those received by the papers published in the first 3 career years;
	        (C) citations received per paper published during career years 4-20;
	        (D) number of different top scientists (per paper published) with whom the researcher has coauthored papers in career years 4-20 (excluding those already accounted in the first 3 career years for the treatment group);
	        (E) number of times (per paper published) the researcher has coauthored papers with a top scientist in career years from 4 to 20 (excluding those already accounted in the first 3 career years for the treatment group).
	        ($^{*}$: $p<0.05$; $^{**}$: $p<0.01$; $^{***}$: $p <0.001$; NS: not significant). Numbers in brackets denote standard errors.} 
	
	\hspace*{-1.3cm}
		\begin{tabular}{c|c|c|c|c|c|c|c|c|c|c|c|c}
		\hline \hline 
		\multirow{2}{*}{}  &  \multicolumn{3}{|c}{Cell Biology} &  \multicolumn{3}{|c}{Chemistry} &  \multicolumn{3}{|c}{Neuroscience} &  \multicolumn{3}{|c}{Physics} \\
		\hline

No. authors		       & \multicolumn{3}{|c}{$2,324$}  & \multicolumn{3}{|c}{$5,635$}       & \multicolumn{3}{|c}{$5,605$} &\multicolumn{3}{|c}{$5,414$}\\
		
No. pairs		       & \multicolumn{3}{|c}{$468$}   & \multicolumn{3}{|c}{$1,443$}    & \multicolumn{3}{|c}{$1,602$} &\multicolumn{3}{|c}{$1,362$}\\
     	\hline 
		&  Top co. & Control & $p$ &  Top co. & Control & $p$ &  Top co. & Control & $p$ &  Top co. & Control & $p$  \\
		\hline 
  	 
Inst. prestige		    & $22.40$  & $22.24$   & NS     & $26.85$ & $26.99$ &  NS  		
             			& $10.52$ & $10.66$ &  NS 		& $19.58$  & $19.69$   & NS     \\
				        & $(0.57)$ & $(0.66)$  &        & $(0.32)$& $(0.39)$&  			
				        & $(0.17)$& $(0.19)$& 			& $(0.24)$ & $(0.27)$  &        \\
				        
Productivity			& $2.36$   & $2.37$    & NS     & $3.09$  & $3.13$  &  NS 		
						& $3.13$  & $3.09$  &  NS 		& $3.35$   & $3.25$    & NS     \\
				        & $(0.06)$ & $(0.06)$  &        & $(0.04)$& $(0.05)$&  			
				        & $(0.04)$& $(0.04)$& 			& $(0.05)$ & $(0.05)$  &        \\
				        				
Cit. (years 1-3)		& $4.80$   & $4.71$    & NS     & $4.06$  & $3.95$  &  NS 		
						& $3.95$  & $3.92$  &  NS 		& $4.22$   & $4.19$    & NS     \\
				        & $(0.15)$ & $(0.14)$  &        & $(0.08)$& $(0.08)$& 			
				        & $(0.07)$& $(0.07)$& 			& $(0.08)$ & $(0.08)$  &        \\
\hline	
\hline 
(A)		&\multicolumn{2}{|c|}{$57.5\%$}& $^{**}$  &\multicolumn{2}{|c|}{$56.3\%$}&$^{***}$  
        &\multicolumn{2}{|c|}{$58.4\%$}&$^{***}$  &\multicolumn{2}{|c|}{$61.2\%$}&$^{***}$\\

(B)	    &\multicolumn{2}{|c|}{$56.6\%$}& $^{**}$  &\multicolumn{2}{|c|}{$56.8\%$}&$^{***}$  
        &\multicolumn{2}{|c|}{$57.9\%$}&$^{**}$   &\multicolumn{2}{|c|}{$61.4\%$}&$^{***}$\\

(C)	    &\multicolumn{2}{|c|}{$48.1\%$} & NS      &\multicolumn{2}{|c|}{$52.3\%$}&$0.08$    
        &\multicolumn{2}{|c|}{$48.4\%$}& NS       &\multicolumn{2}{|c|}{$56.0\%$} &$^{***}$\\

(D)	    &\multicolumn{2}{|c|}{$50.6\%$}  & NS     &\multicolumn{2}{|c|}{$57.2\%$}&$^{***}$  
        &\multicolumn{2}{|c|}{$52.9\%$}&$^{*}$    &\multicolumn{2}{|c|}{$56.8\%$}  &$^{***}$\\

(E)	    &\multicolumn{2}{|c|}{$52.6\%$}  & NS     &\multicolumn{2}{|c|}{$59.7\%$}&$^{***}$  
        &\multicolumn{2}{|c|}{$54.7\%$}&$^{***}$  &\multicolumn{2}{|c|}{$59.2\%$}  &$^{***}$\\

\hline \hline 
\end{tabular}
\label{table:match_app}
\end{table}

\begin{figure*}[ht!]

	\centering
	\includegraphics[width=.75\linewidth]{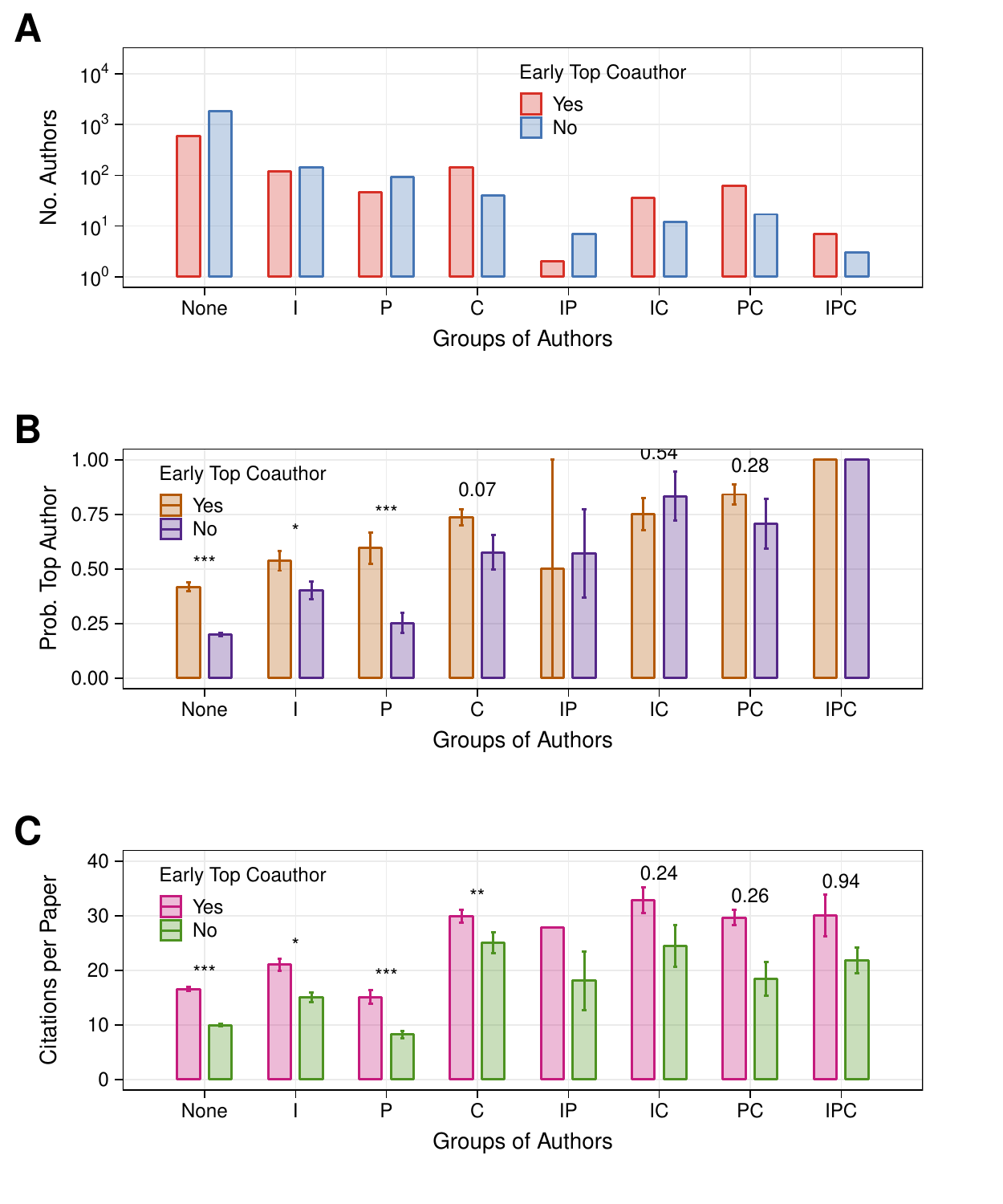}
	\caption{\textbf{Relationship between long term impact and early career performance in Cell Biology.} \textbf{A} Number of junior researchers belonging to the top 10\% in various categories of early career performance (I denotes institutional prestige, P denotes productivity, C denotes citations received. All three such quantities are computed based on the first 3 career years). \textbf{B} Probability for authors belonging to each group of being a top scientist in their 20th career year. \textbf{C} Number of citations received per paper published by authors belonging to each group between their 4th and 20th career year. In all panels, we report the $p$-values obtained via $t$-tests to assess the statistical significance of differences between the sub-group of junior researchers who coauthor work with a top scientist in the first 3 career years and the sub-group of those who do not. $^{*}$: $p<0.05$; $^{**}$: $p<0.01$; $^{***}$: $p <0.001$}
	\label{fig:cell_app}
\end{figure*}
\newpage
\begin{figure*}[ht!]
	\centering
	\includegraphics[width=.75\linewidth]{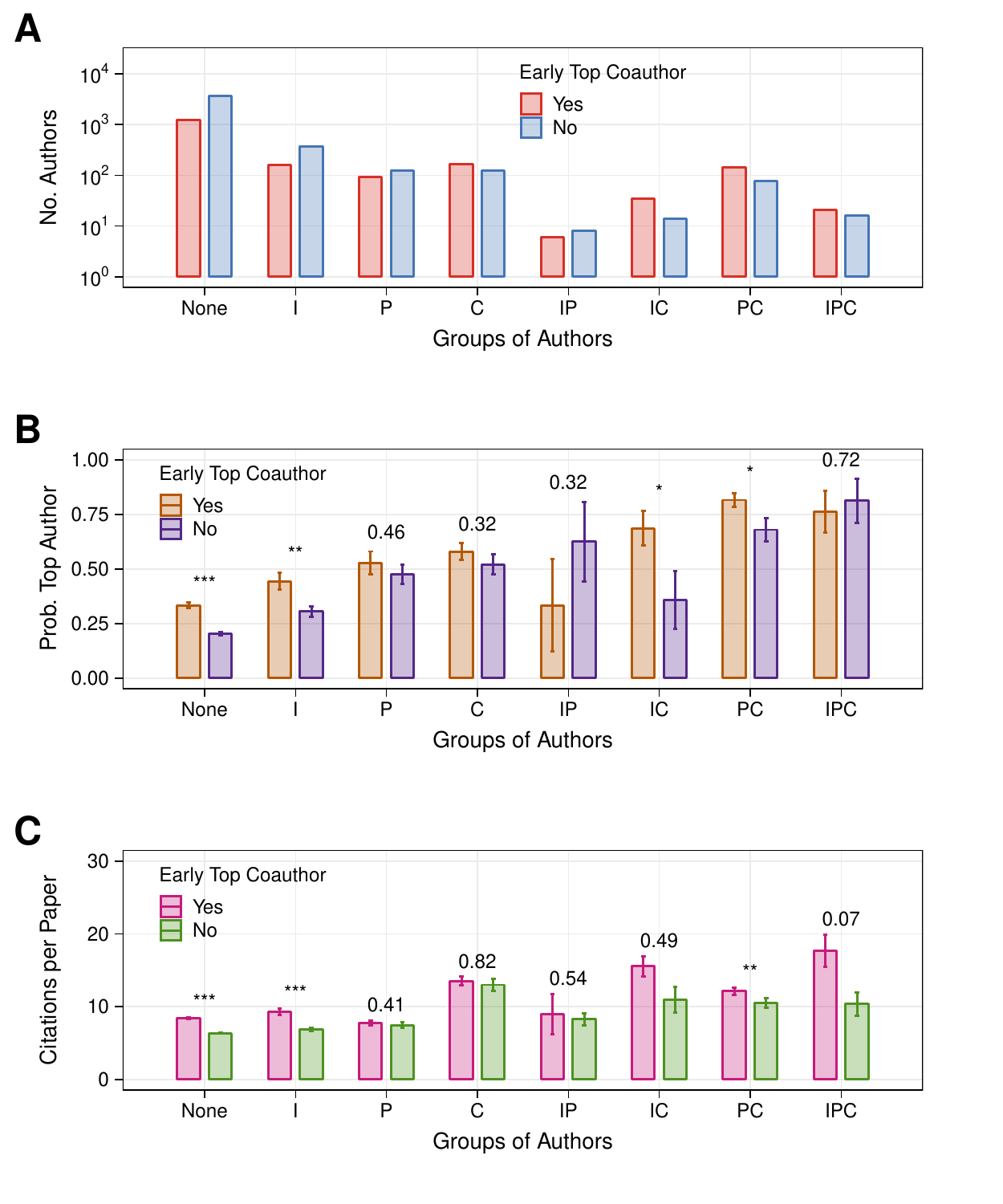}
	\caption{\textbf{Relationship between long term impact and early career performance in Chemistry.} \textbf{A} Number of junior researchers belonging to the top 10\% in various categories of early career performance (I denotes institutional prestige, P denotes productivity, C denotes citations received. All three such quantities are computed based on the first 3 career years). \textbf{B} Probability for authors belonging to each group of being a top scientist in their 20th career year. \textbf{C} Number of citations received per paper published by authors belonging to each group between their 4th and 20th career year. In all panels, we report the $p$-values obtained via $t$-tests to assess the statistical significance of differences between the sub-group of junior researchers who coauthor work with a top scientist in the first 3 career years and the sub-group of those who do not. $^{*}$: $p<0.05$; $^{**}$: $p<0.01$; $^{***}$: $p <0.001$}
	\label{fig:chemistry_app}
\end{figure*}
\newpage
\begin{figure*}[h!]
\vspace{-1cm}
	\centering
	\includegraphics[width=.75\linewidth]{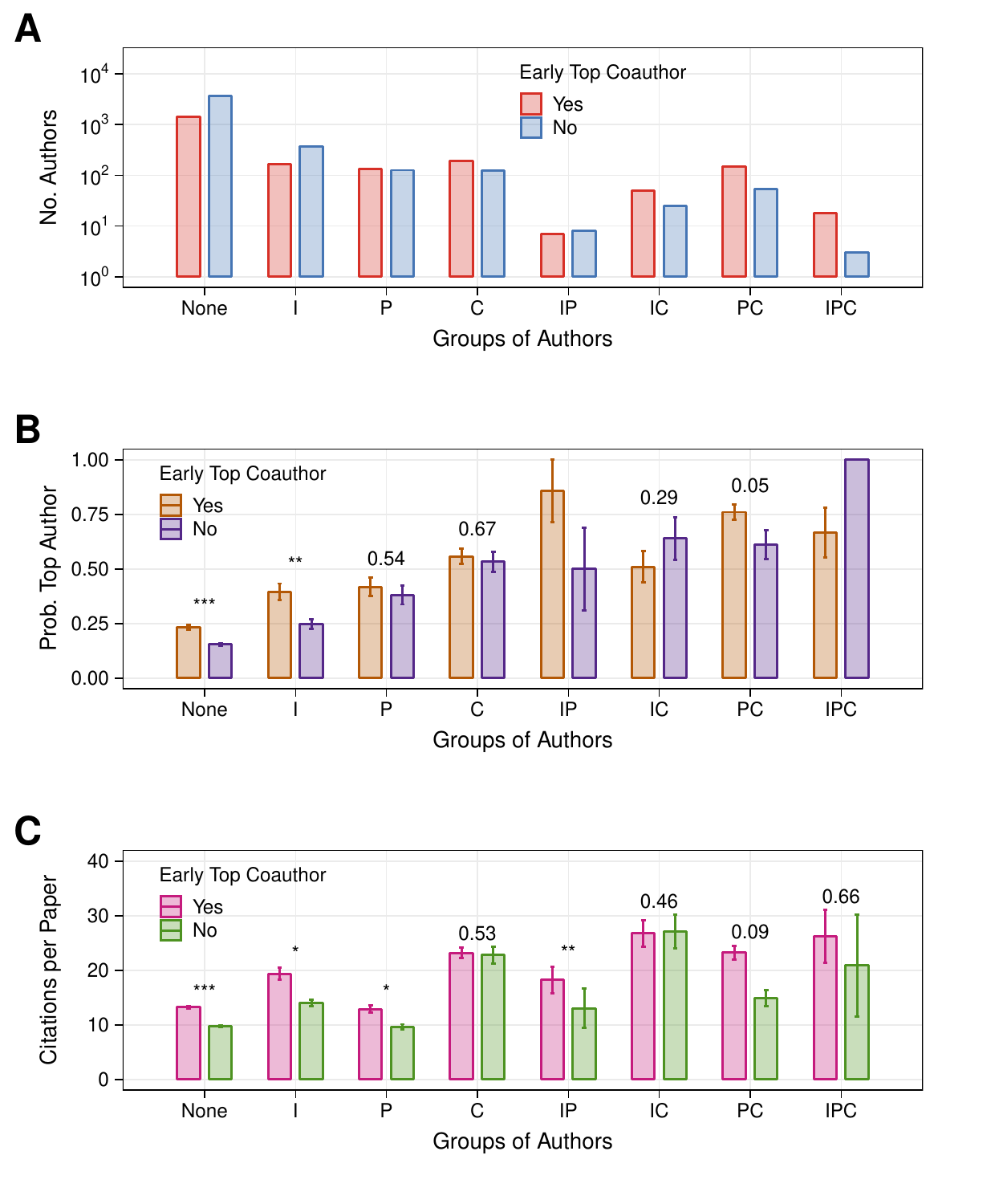}
	\caption{\textbf{Relationship between long term impact and early career performance in Neuroscience.} \textbf{A} Number of junior researchers belonging to the top 10\% in various categories of early career performance (I denotes institutional prestige, P denotes productivity, C denotes citations received. All three such quantities are computed based on the first 3 career years). \textbf{B} Probability for authors belonging to each group of being a top scientist in their 20th career year. \textbf{C} Number of citations received per paper published by authors belonging to each group between their 4th and 20th career year. In all panels, we report the $p$-values obtained via $t$-tests to assess the statistical significance of differences between the sub-group of junior researchers who coauthor work with a top scientist in the first 3 career years and the sub-group of those who do not. $^{*}$: $p<0.05$; $^{**}$: $p<0.01$; $^{***}$: $p <0.001$}
	\label{fig:neuroscience_app}
\end{figure*}

\newpage
\begin{figure*}[ht!]
\vspace{-1cm}
	\centering
	\includegraphics[width=.75\linewidth]{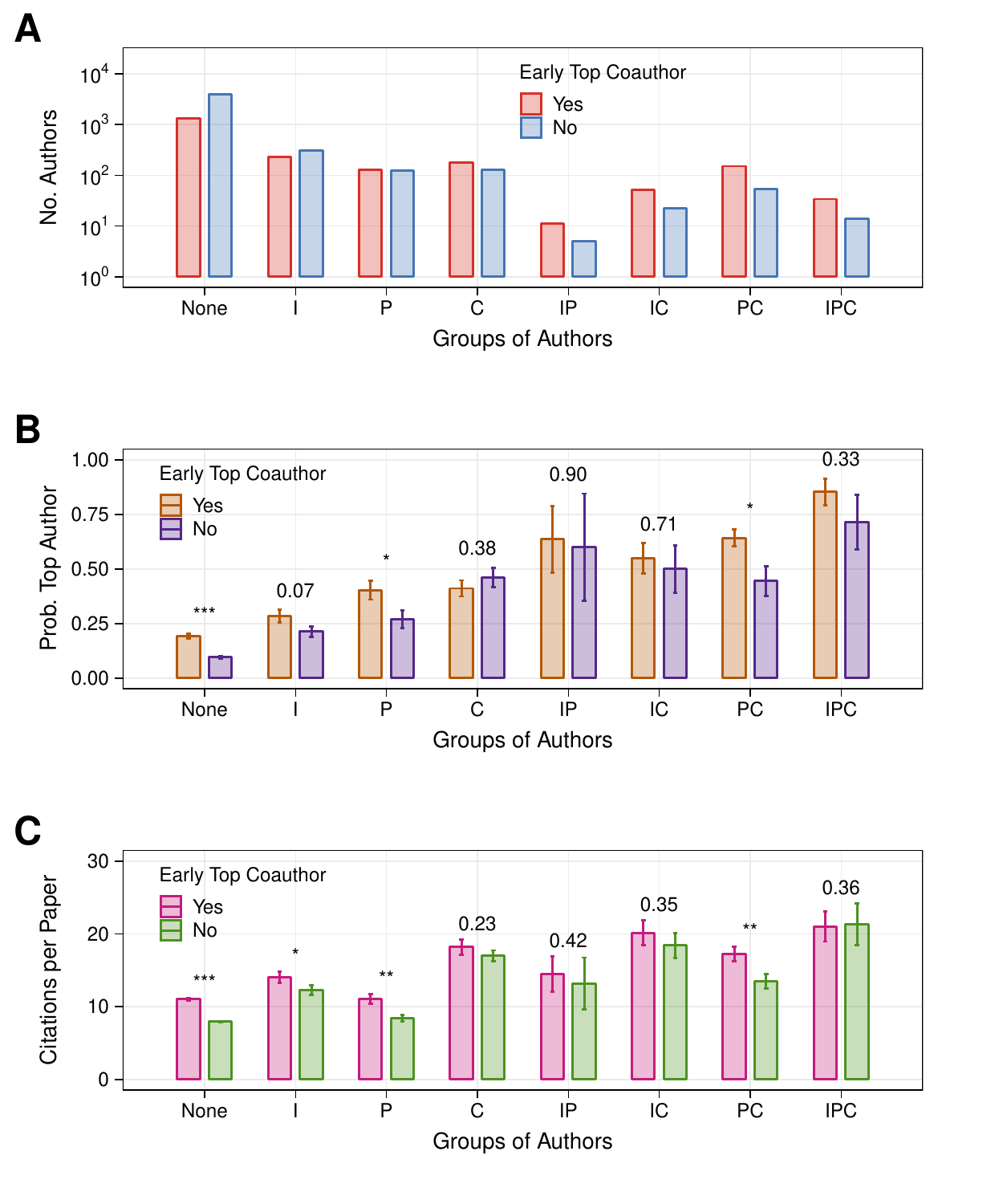}
	\caption{\textbf{Relationship between long term impact and early career performance in Physics.} \textbf{A} Number of junior researchers belonging to the top 10\% in various categories of early career performance (I denotes institutional prestige, P denotes productivity, C denotes citations received. All three such quantities are computed based on the first 3 career years). \textbf{B} Probability for authors belonging to each group of being a top scientist in their 20th career year. \textbf{C} Number of citations received per paper published by authors belonging to each group between their 4th and 20th career year. In all panels, we report the $p$-values obtained via $t$-tests to assess the statistical significance of differences between the sub-group of junior researchers who coauthor work with a top scientist in the first 3 career years and the sub-group of those who do not. $^{*}$: $p<0.05$; $^{**}$: $p<0.01$; $^{***}$: $p <0.001$}
	\label{fig:physics_app}
\end{figure*}

\end{document}